\documentclass[a4paper,12pt]{article}
\pdfoutput=1 

\usepackage{jcappub} 
\usepackage{subfigure}
\usepackage[T1]{fontenc} 

\def\hmath$#1${\texorpdfstring{{\rmfamily\textit{#1}}}{#1}}
\usepackage{hyperref}
\usepackage{amsfonts}
\usepackage{threeparttable}

\subheader{IIPDM-2019}

\title{\boldmath Search for Semi-Annihilating Dark Matter with Fermi-LAT, H.E.S.S., Planck, and the Cherenkov Telescope Array}

\author[a]{Farinaldo S. Queiroz,}
\author[a]{Clarissa Siqueira}

\affiliation[a]{International Institute of Physics, Universidade Federal do Rio Grande do Norte,
Campus Universitario, Lagoa Nova, Natal-RN 59078-970, Brazil}

\emailAdd{farinaldo.queiroz@iip.ufrn.br}
\emailAdd{csiqueira@iip.ufrn.br}

\abstract{
Dark matter annihilations have been fiercely restricted by the non-observation of excess events in indirect detection probes. Typically the interactions that dictate annihilation are also present in the dark matter-nucleon scattering cross section, also severely constrained by direct detection experiments. Semi-annihilations arise as a possible way to ameliorate the tension and even change the standard dark matter relic density calculation. In this work, we derive indirect detection bounds for several semi-annihilation channels including gauge bosons, the Higgs, leptophilic and leptophobic scalars. Our analysis is based on the gamma-ray observations in the direction of Dwarf Spheroidal Galaxies (Fermi-LAT) and the Galactic Center (H.E.S.S.), and Planck measurements of Cosmic Background Radiation. In addition, we derive the prospects for the Cherenkov Telescope Array (CTA) sensitivity to all these semi-annihilation modes.
}

\begin{document}

\maketitle
\flushbottom

\section{Introduction}
\label{sec:intro}

The presence of an abundant non-baryonic dark matter component in our universe is one of the most exciting phenomena in nature \cite{Queiroz:2016sxf}. Its presence has been established through a variety of independent observations \cite{Peter:2012rz,Gaskins:2016cha,Kahlhoefer:2017dnp}. Dark matter is key to the evolution of our universe and the formation of galaxies as we know it \cite{Primack:1997av}. Therefore, it is important to unveil its nature. Basically, all astrophysical attempts to account for dark matter in our universe have failed, leaving one compelling option behind, an elementary particle.\\

Interpreting dark matter in terms of elementary particles requires physics beyond the Standard Model. The Standard Model (SM) does not have any particle that could fulfill all the requirements to play the role of dark matter. Several dark matter candidates have been proposed \cite{Feng:2010gw}, and the most theoretically appealing are the WIMPs (Weakly Interacting Massive Particles) \cite{Arcadi:2017kky}. Dark matter models that feature WIMPs suffer from severe bounds from direct and indirect detection experiments \cite{Angle:2008we,Aguilar-Arevalo:2016ndq,Fermi-LAT:2016uux,Ahnen:2016qkx,Amole:2017dex,TheFermi-LAT:2017vmf,Cui:2017nnn,Akerib:2017kat,Agnes:2018ves}. In vanilla dark matter models, dark matter abundance is set by the dark matter annihilation cross section, where the WIMP self-annihilates producing SM particles, $\chi \chi \rightarrow SM\, SM$ \cite{Arcadi:2017kky}. Typically the parameters that govern the dark matter self-annihilation also dictate the dark matter-nucleon scattering cross section, which is severely restricted by direct detection experiments. It is worth noting that one easy way to weaken the direct detection bounds is by increasing the dark matter mass since the direct detection bounds are proportional to the number density of dark matter particles. \\

Different methods have surfaced to weaken the relation between the dark matter annihilation cross section and the scattering cross section. Models that feature large coannihilations \cite{Arcadi:2017xbo}, secluded annihilations \cite{Profumo:2017obk,Campos:2017odj,Dutra:2018gmv}, semi-annihilations \cite{Hambye:2008bq,Arina:2009uq,Hambye:2009fg,DEramo:2010keq,Belanger:2012vp,DEramo:2012fou,Aoki:2014cja,Arcadi:2017vis,Kamada:2017gfc,Kamada:2018hte}, and non-thermal production \cite{Bernal:2017kxu,Bernal:2018hjm} are among the most popular mechanism to untie direct detection observables to the dark matter relic density.\\

In this work, we will focus on semi-annihilations. In a dark matter model, there might be several particles that live in the dark sector. Differently from coannihilation and annihilation processes, where only SM particles are present in the final state, semi-annihilations refer to annihilations where three particles from the dark sector appear in the final state. This alters the Boltzmann equation which controls the overall dark matter abundance \cite{Cai:2018imb}. Moreover, in some cases, there can be basically no direct detection signals, explaining the absence of solid positive signals in direct detection experiments. This strengthens the role of indirect detection probes.\\

Indirect detection experiments such as Fermi-LAT, H.E.S.S., among others, have not provided their sensitivity semi-annihilations. Motivated by the theoretical relevance of semi-annihilations and the absence of indirect detection limits on them, we derive in this work bounds on the semi-annihilation cross section for a variety of setups. In particular, we use the gamma-ray observation from Fermi-LAT and H.E.S.S. telescopes in the direction of Dwarf Spheroidal Galaxies (dSphs) and the Galactic Center, respectively. Furthermore, we use the Planck satellite measurements on the Cosmic Microwave Background (CMB) to constrain semi-annihilations processes. Thus, we use two completely different datasets to constrain semi-annihilations. Lastly, we obtain the Cherenkov Telescope Array (CTA) sensitivity to semi-annihilations having in mind that CTA will be the most sensitive gamma-ray telescope to dark matter masses below $10$~TeV.\\

Our work is structured as follows: In Section \ref{sec:dmann} we describe the type of semi-annihilations we will investigate; in section \ref{sec:indirect} we discuss the datasets; in section \ref{sec:results} we present our results and later conclude.
 
\section{Semi-Annihilation Dark Matter}
\label{sec:dmann}

Dark matter particles should be stable at cosmological scales, with a lifetime much larger than the age of the universe \cite{Baring:2015sza,Fornasa:2016ohl,Ando:2016ang,Hoof:2018hyn,Mitridate:2018iag,Clark:2018ghm}. The stability of the dark matter particle is usually guaranteed by a discrete $Z_2$ symmetry. In such models semi-annihilations are absent, and indirect detection bounds on the dark matter self-annihilation cross section have been placed \cite{Hooper:2012sr,Li:2013qya,Li:2015kag,Giesen:2015ufa}. However, if larger symmetries are behind the dark matter stability, semi-annihilations can be very important as it happens in models with $Z_3$, $Z_4$ symmetries and in more complex dark sectors~\cite{Belanger:2012zr,Guo:2015lxa,Ko:2014loa,Belanger:2014bga,Cai:2015tam,Cai:2015zza,Cai:2016hne,Camargo:2019ukv,Hektor:2019ote}.\\

Generally speaking semi-annihilations refer to processes of type $\chi_1 + \chi_2 \to \chi_3 +\mathrm{X}$, where X can either be a SM particle or not.\\

In principle $\chi_1,\chi_2,\chi_3$ can represent different particles. It is well-known that in case which $\chi_1$ is  different from $\chi_{2,3}$, there might be co-annihilations at which change the dark matter relic density if the mass splitting is small. However, for simplicity, we will assume that $\chi_1$, $\chi_2$ and $\chi_3$ represent only one particle, the dark matter, $DM$. From now on will use the label $DM$ to refer to the dark matter particle.\\


That said, the accompanying particle, $X$, can be either a scalar, a fermion or a vector. The dark matter mass and possible decay channels for $X$ will determine which indirect detection experiment is more sensitive. For this reason, our analyses will be based on different data-sets and experiments.\\

There are several semi-annihilation setups to be investigated. In this work, we will address some of them, in some cases involving a SM particle in the final state, other involving new scalars with particular decay modes. In the first one, the final state particle is the SM Z or the Higgs boson, $h$, the branching ratios are fixed. On the other one, it is possible to choose a non-standard particle in the final state with different decay patterns and masses.\\

In order to encompass several possibilities in our analyses will investigate:
\begin{itemize}
    \item  (i) $DM + DM \rightarrow DM+ Z$;
    \item (ii) $DM + DM \rightarrow DM + h$, where $h$ is the SM Higgs boson;
    \item (iii) $DM + DM \rightarrow DM + \phi$, where $\phi$ is Higgs-like particle but with different mass, $m_\phi=10$~GeV, $100$~GeV, $500$~GeV;
    \item (iv) $DM + DM\, \rightarrow DM + \phi$, where $\phi$ is a leptophilic scalar decaying exclusively into $e^+e^-$, $\mu^+\mu^-$, $\tau^+\tau^-$;
     \item (v) $DM + DM \rightarrow DM + \phi$, where $\phi$ is a leptophobic scalar decaying into $b\bar{b}$ with a $100\%$ branching ratio.
\end{itemize}

In what follows we will describe the datasets used in our analyses.

\section{Indirect Detection} 
\label{sec:indirect}

\subsection{Gamma-rays}
\label{sec:fermi}

Dark matter indirect detection constitutes an interesting way to probe particle dark matter models. Targets with high-density regions as the Galactic Center and Dwarf Spheroidal Galaxies (dSphs) represent good targets. Dark matter annihilations and semi-annihilations into electrons, positrons, quarks, neutrinos etc., yield lots of gamma-rays with different energy spectra. Therefore, gamma-ray observations are of the foremost importance. If we understand well the gamma-ray emission from astrophysical objects, we can use this information to probe the presence of dark matter annihilations and semi-annihilations by comparing the expected gamma-ray flux to the measured one since any signal from dark matter interactions will appear as a surplus of gamma-ray emission.\\

The expected gamma-ray flux coming from dark matter annihilation, $DM + DM \rightarrow SM + SM$, where $DM$ represents the dark matter particle and $SM$ the standard model ones, is given by,

\begin{equation}
 \frac{d\phi_\gamma}{dE}=\frac{1}{8\pi m_{DM}^2} \langle \sigma v \rangle  \frac{dN_\gamma}{dE} J
 \label{semiflux}
\end{equation}
where $m_{DM}$ is the DM particle mass, $ \langle \sigma v \rangle $ is the annihilation cross section, $dN_\gamma/dE$ the gamma-ray spectrum, and $J$ is the J-factor which takes into account all the astrophysical part of the process, 

\begin{equation}
 J=\int_{l.o.s}\frac{ds}{r_\odot}\left(\frac{\rho(r(s,\theta))}{\rho_\odot}\right)^2
\end{equation}which included an integration over the line-of-sight between the observatory and the source. This factor depends on the dark matter distribution, which we adopted to follow an Einasto profile,

\begin{equation}
    \rho_{Ein}(r) = \rho_{s} \exp{ \left( \frac{-2}{\alpha}\left[\left(\frac{r}{r_{s}}\right)^{\alpha}-1 \right] \right)},
\end{equation}
where $r$ is galactic radius, $r_s=28.44$~kpc is a typical scale radius, $\rho_s=0.033$ is a typical scale density and, $\alpha=0.17$ \cite{Cirelli:2010xx}, that assumes a spherical dark matter distribution in the halo. \\

As aforementioned, in this work, instead of concentrating on the standard dark matter annihilation setup, $DM + DM \rightarrow SM + SM$, we focus on semi-annihilations, $DM + DM \rightarrow DM + X$, where $X$ is a neutral particle may belong to the SM spectrum or not. The key difference for the flux computation in Eq.~\ref{semiflux} for the semi-annihilation process goes into the energy spectrum ($dN/dE$) since this alters the amount of photons allowed for each energy bin. In order to compute the gamma-ray spectrum $dN/dE$, we use the numerical package Pythia8 \cite{Sjostrand:2014zea}. There other spectrum generators in the literature \cite{Cirelli:2010xx}, but they do not allow for the possibility to incorporate the channels sifted here.


With these ingredients at hand, for different choices of the final state annihilation channel, dark matter mass, and semi-annihilation cross section we can compute the gamma-ray flux produced by dark matter interactions to be compared with the experimental data. As far as gamma-rays are concerned we investigate three different datasets:

\begin{itemize}
    \item Fermi-LAT gamma-ray observations in the direction of dwarf spheroidal galaxies;
    \item H.E.S.S. gamma-ray observations in the direction of the Galactic Center;
    \item The Cherenkov Telescope Array projected sensitivity to gamma-rays stemming from the Galactic Center.
\end{itemize}

We describe these datasets in more detail below. \\

The {\it Fermi-LAT satellite} made one of the most remarkable mappings of the sky in gamma rays, covering an energy range between $500$~MeV and $500$~GeV \cite{Ackermann:2012nb,Charles:2016pgz}. After many years of observations no solid excess has been observed in gamma-rays in the direction of dSphs, and consequently, stringent limits were imposed on the dark matter annihilation cross section for the standard annihilation channels \cite{Ackermann:2015zua}, namely $b\bar{b}$, $\tau\bar{\tau}$, etc. In this work, we use 6 years of data from 15 dSphs (see Table \ref{tab:dsphs}), based on the \textbf{pass 8} event analysis \cite{Ackermann:2015zua} to compute the exclusion limits for dSphs in semi-annihilation models. In order to derive robust limits taking into account the error in the J-factors, as shown in {\it Table} \ref{tableJfactor}, we used the likelihood function for a combined analysis of 15 dSphs using the package provided in \cite{Workgroup:2017lvb}. If the statistical errors in the J-factors were ignored, more optimistic bounds would have been found.  

\begin{table}[h!]
\centering
\begin{threeparttable}
\caption{ \label{tab:dsphs} Milky way dSphs' properties \cite{Ackermann:2015zua}}
\begin{tabular}{ |l |r |r| c |c |l |}
\hline
  Name                      & $\ell$\tnote{1} & $b$\tnote{1} & Dist. & $\log_{10}({\ensuremath{J_{\textrm{obs}}}})$\tnote{2} &  Ref. \\
  & (deg) & (deg) & (kpc) & {\small ($\log_{10}(\text{GeV}^2 \text{cm}^{-5})$)} &   \\
  \hline
  Bootes I                  & 358.1  & 69.6   & 66     & $18.8 \pm 0.22$ & \cite{DallOra:2006tki} \\
  Canes Ven. II         & 113.6  & 82.7   & 160    & $17.9 \pm 0.25$ & \cite{Simon:2007dq} \\
  Carina                    & 260.1  & $-$22.2  & 105    & $18.1 \pm 0.23$ & \cite{Walker:2008ax} \\
  Coma Ber.            & 241.9  & 83.6   & 44     & $19.0 \pm 0.25$ & \cite{Simon:2007dq} \\
  Draco                     & 86.4   & 34.7   & 76     & $18.8 \pm 0.16$ & \cite{Munoz:2005be} \\
  Fornax                    & 237.1  & $-$65.7  & 147    & $18.2 \pm 0.21$ & \cite{Walker:2008ax} \\
  Hercules                  & 28.7   & 36.9   & 132    & $18.1 \pm 0.25$ & \cite{Simon:2007dq} \\
  Leo II                    & 220.2  & 67.2   & 233    & $17.6 \pm 0.18$ & \cite{Koch:2007ye} \\
  Leo IV                    & 265.4  & 56.5   & 154    & $17.9 \pm 0.28$ & \cite{Simon:2007dq} \\
  Sculptor                  & 287.5  & $-$83.2  & 86     & $18.6 \pm 0.18$ & \cite{Walker:2008ax} \\
  Segue 1                   & 220.5  & 50.4   & 23     & $19.5 \pm 0.29$ & \cite{Simon:2010ek} \\
  Sextans                   & 243.5  & 42.3   & 86     & $18.4 \pm 0.27$ & \cite{Walker:2008ax} \\
  Ursa Maj. II             & 152.5  & 37.4   & 32     & $19.3 \pm 0.28$ & \cite{Simon:2007dq} \\
  Ursa Minor                & 105.0  & 44.8   & 76     & $18.8 \pm 0.19$ & \cite{Munoz:2005be} \\
  Willman 1                 & 158.6  & 56.8   & 38     & $19.1 \pm 0.31$ & \cite{Willman:2010gy} \\
  \hline
\end{tabular}
\label{tableJfactor}
\begin{tablenotes}
\item[1] Galactic longitude and latitude.
\item[2] The J-factors were computed using NFW profile.
\end{tablenotes}
\end{threeparttable}
\end{table}

The \textit{High Energy Stereoscopic System} (H.E.S.S.) gave us a new look at high energy gamma-rays. Its location in Namibia is privileged to look at the Galactic Center. It is currently the most sensitive gamma-ray detector to dark matter masses of about $100$~TeV. We use the H.E.S.S. II \cite{Abdallah:2016ygi} data\footnote{Following the prescription described in \cite{secluded} since the collaboration does not share the data, we compute the results for the H.E.S.S. I, using the package \cite{Workgroup:2017lvb} and just re-scale the result, which is good agreement with H.E.S.S. analysis.} in a region of $1^\circ$ surrounding the Galactic Center, excluding the region  $|b|<0.3^\circ$, due to the high background. Using this data we use again the package \cite{Workgroup:2017lvb} to compute the likelihood function following the description in \cite{Abramowski:2011hc}. \\

The {\textit Cherenkov Telescope Array} (CTA) is an experiment projected to detect high-energy gamma-rays. In order to cover a wide energy range ($\sim 60$~GeV to $\sim 300$~TeV), it is composed by telescopes with different sizes, including small, medium and large-scale telescopes, located at Chile (Southern site) and La Palma (Northern site), in order to cover the entire sky. Based on \cite{Silverwood:2014yza}, we use the combined morphological analysis of the Galactic Center gamma-ray emission in order to compute the projected limits on the dark matter semi-annihilation cross section for several channels. To the best of our knowledge there no sensitivity reach of the CTA to semi-annihilation channels. Using the recipe discussed in the previous section to account for semi-annihilations, we computed the likelihood functions for CTA  and we were able to compute the limits on the semi-annihilation cross-section through the statistical test (TS), 

\begin{equation}
    TS = -2 \ln{\left( \frac{\mathcal{L}(\hat{\mu}_0,\hat{\theta}|\mathcal{D})}{\mathcal{L}(\hat{\mu},\hat{\theta}|\mathcal{D})}  \right)},
\end{equation}
where taking $TS>2.71$ we get the $95\%$~C.L. exclusion curve. The $\hat{\mu}$ and $\hat{\mu}_0$ parameters are related to the best-fit model parameters with dark matter and to the null dark matter hypothesis, respectively, $\hat{\theta}$ the parameters from the background, and $\mathcal{D}$ is the experimental data.\\  

After considering three different datasets based on gamma-ray observations, we will outline below an orthogonal probe to semi-annihilation which rely on the precise measurements of the Cosmic Background Radiation.

\subsection{Cosmic Background Radiation}
\label{sec:planck}

Products of Dark Matter annihilation during the period of recombination can affect the Cosmic Microwave Background (CMB) anisotropies due to electromagnetic energy injection in the primordial plasma \cite{Slatyer:2015jla}. This energy injection is given by,
\begin{equation}
\frac{dE}{dtdV}=\rho_c^2 c^2 \Omega_{DM}^2 (1+z)^6 P_{ann}(z)
\end{equation}
where $\rho_c$ is the critical density, $c$ is the light velocity, $\Omega_{DM}$ is the DM abundance, and $z$ is the redshift, the annihilation parameter ($P_{ann}$) is found to be,
\begin{equation}
P_{ann}=f_{eff}\frac{\langle \sigma v \rangle}{m_{DM}}.
\label{EqPann}
\end{equation}

\begin{figure}[ht]
\centering
\includegraphics[width=0.49\columnwidth]{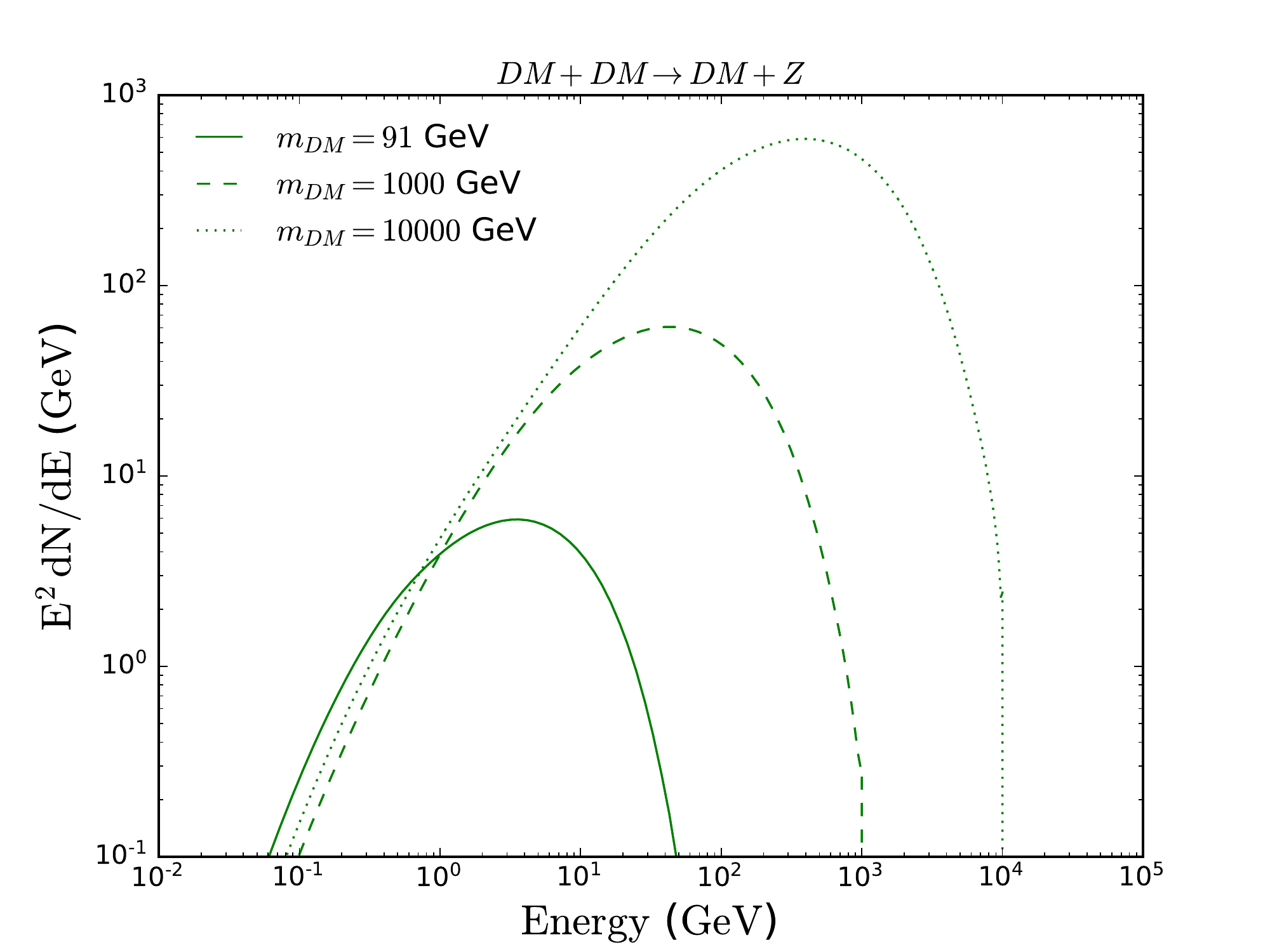}
\includegraphics[width=0.49\columnwidth]{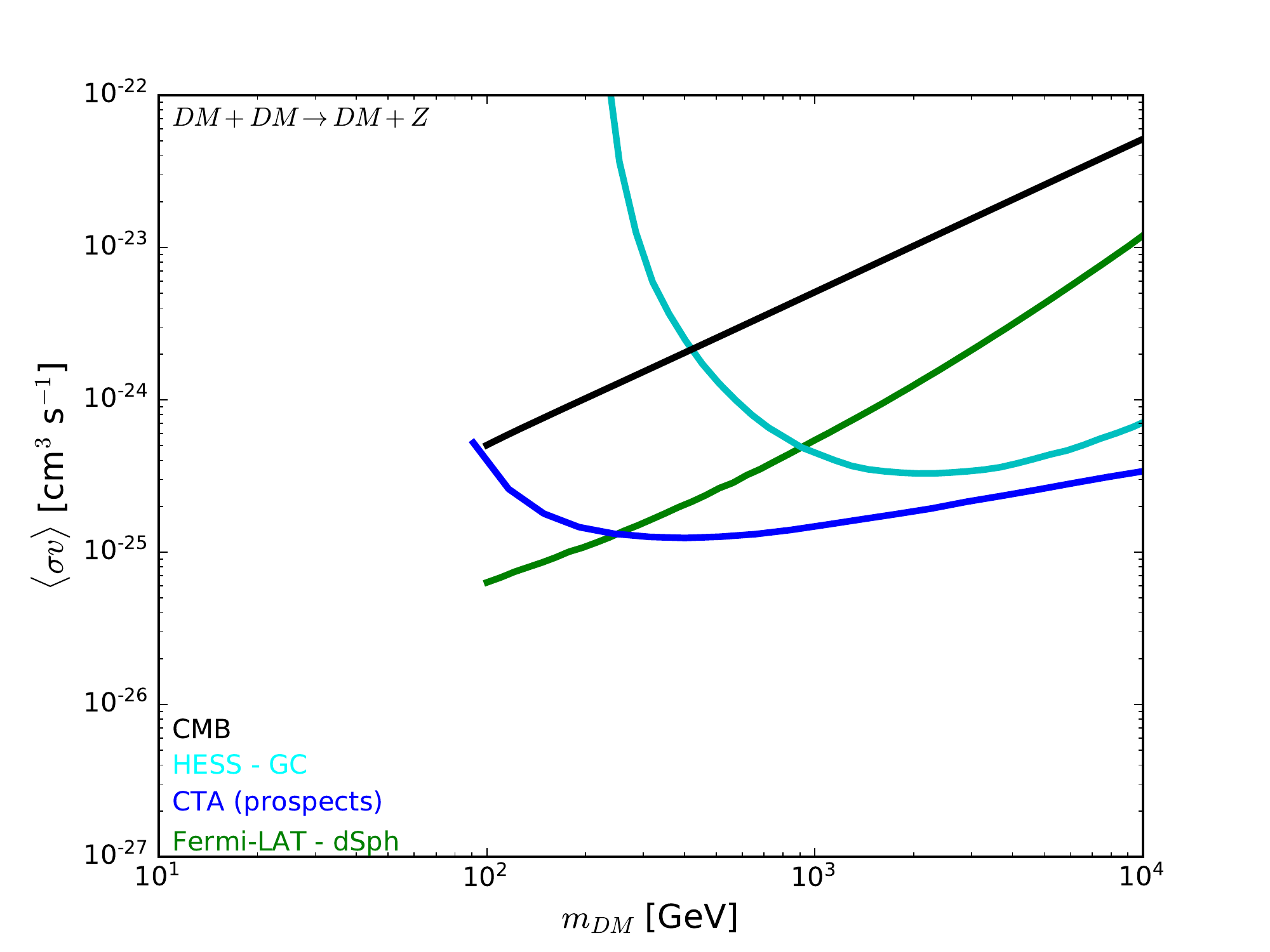}
\caption{\textbf{Left-panel}: Gamma-ray spectra for dark matter semi-annihilating into the SM Z boson, $DM +DM\rightarrow DM+Z$, for dark matter masses equal to $91$~GeV (continuous line), $1000$~GeV (dashed line) and $10000$~GeV (dotted line). \textbf{Right-panel}: Limits from different experiments over the semi-annihilation cross-section. Black lines for Planck constraints, cyan lines for the H.E.S.S. experiment, green curves for the Fermi-LAT experiment and blue lines for the CTA prospect.}
\label{semiz}
\end{figure}

This energy injection can delay the last scattering surface preventing the formation of the first atoms. In other words, by precisely measuring the CMB power spectrum we can probe particle physics effects that took place in the dark age. \\

The efficiency function, $f_{eff}$ in Eq.~\ref{EqPann}, relates the injected and the deposited energy in the thermal bath, the semi-annihilation cross section $\langle \sigma v \rangle$ and the dark matter mass $m_{DM}$. The efficiency function can be parametrized as a function of the efficiency function for photons ($f_{eff}^{\gamma}$), electrons ($f_{eff}^{e^-}$) and positrons ($f_{eff}^{e^+}$) times the associated spectra ($dN/dE^{\gamma,e^+,e^-}$),
\begin{equation}
f_{eff}=\frac{1}{2m_{DM}}\int_0^{m_{DM}}EdE\left(f_{eff}^{\gamma}(E)\frac{dN}{dE^{\gamma}}+2 f_{eff}^{e^+}(E)\frac{dN}{dE^{e^+}}\right).
\end{equation}

Using the constraint on the annihilation parameter imposed by the Planck experiment \cite{Ade:2015xua},
\begin{equation}
P_{ann} < 4.1 \times 10^{-28} \ \ \mathrm{cm}^3 \, \mathrm{s}^{-1} \, \mathrm{GeV}^{-1} 
\end{equation} we compute the limits over the semi-annihilation cross section for a wide range of channels. To do so, we calculate numerically the efficiency function using the routine available in \cite{Slatyer:2015jla}, and we use the Pythia 8 package \cite{Sjostrand:2014zea} to compute the energy spectrum. \\

Now we have summarized the datasets relevant to our reasoning will exhibit the limits below.

\section{Results}
\label{sec:results}

\par In this section, we present our results where we set limits for the first time on the dark matter semi-annihilation cross section for several channels using experimental data from Fermi-LAT, H.E.S.S., Planck, and the prospects for the CTA consortium. We consider different channels, namely: (i) $DM + DM \rightarrow DM + h$; (ii) $DM + DM \rightarrow DM + Z$; (iii) $DM + DM \rightarrow DM + \phi$, where $\phi$ is a Higgs-like scalar with different mass, $DM + DM \rightarrow DM + \phi$, where $\phi$ is a leptophilic scalar, $DM + DM \rightarrow DM + \phi$, where $\phi$ is leptophobic. 

\begin{figure}[ht]
\centering
\includegraphics[width=0.49\columnwidth]{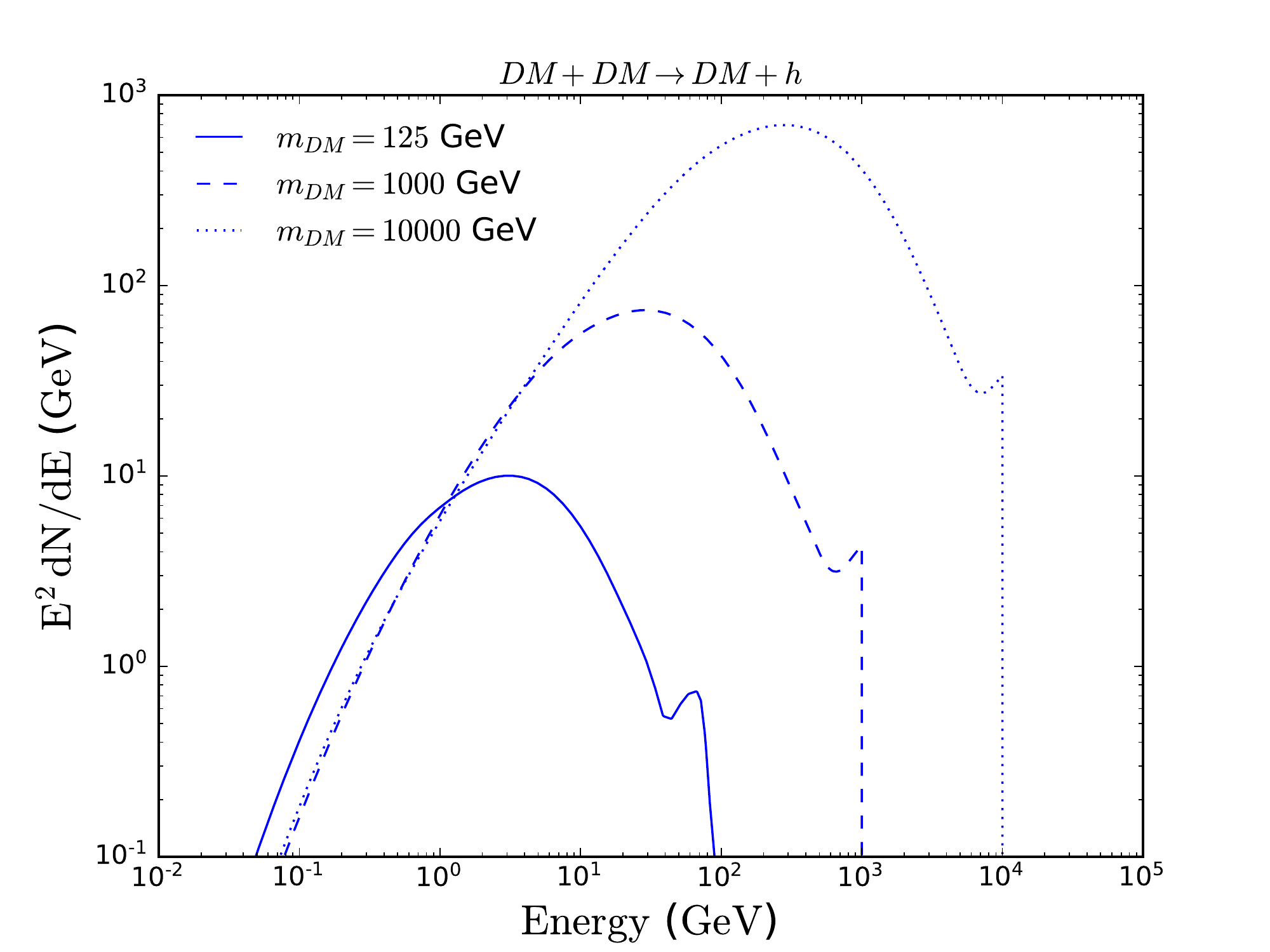}
\includegraphics[width=0.49\columnwidth]{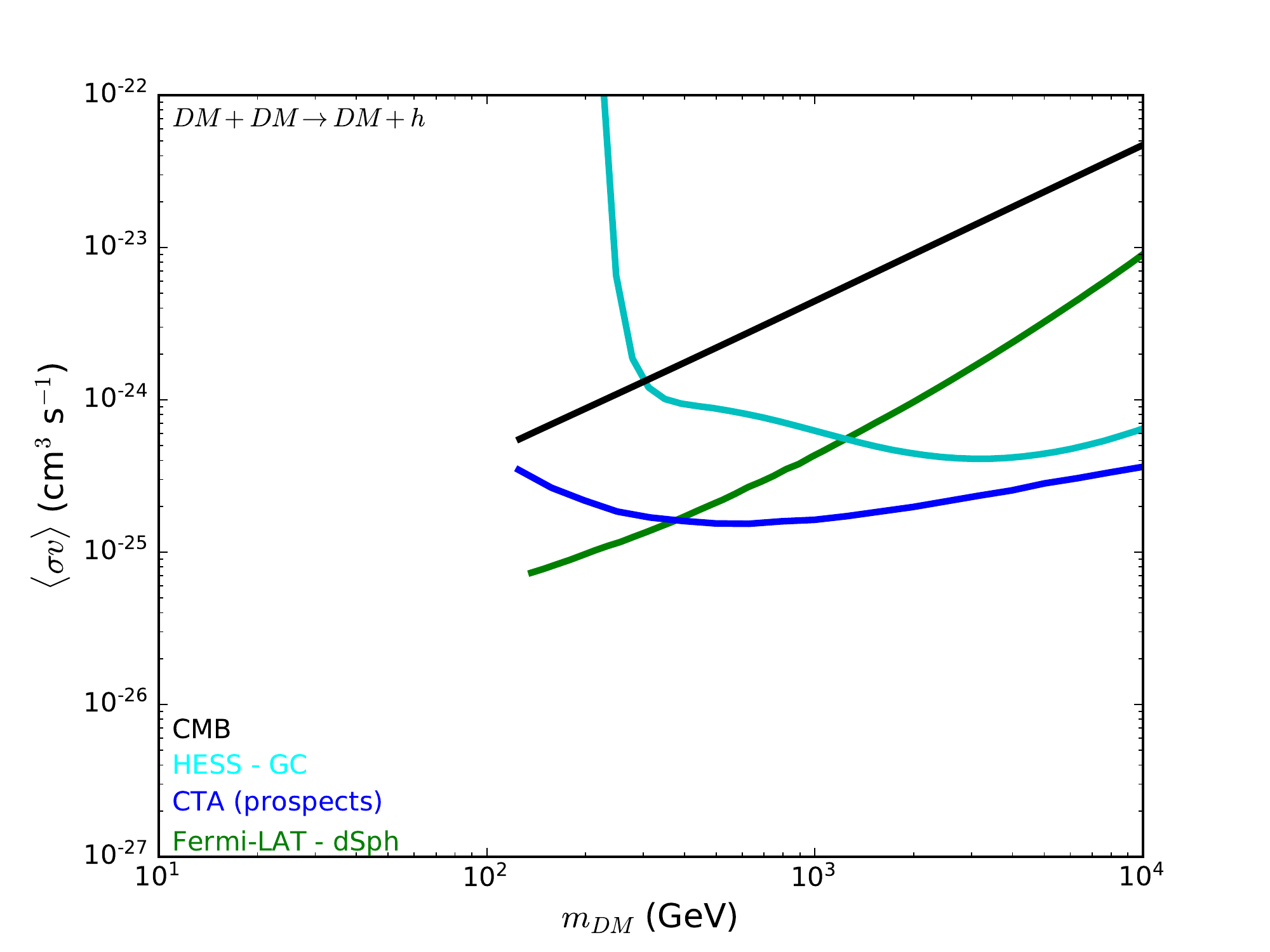}
\caption{\textbf{Left-panel}: Gamma-ray spectrum for dark matter semi-annihilating in to Higgs, $DM+DM \rightarrow DM+h$, for dark matter masses of $125$~GeV (continuous line), $1000$~GeV (dashed line) and $10000$~GeV (dotted line). \textbf{Right-panel}: Limits from different experiments over the semi-annihilation cross section. Black lines for Planck constraints, cyan lines for the H.E.S.S. experiment, green curves for the Fermi-LAT experiment and blue lines for the CTA prospect.}
\label{semih}
\end{figure}

\subsection{Semi-Annihilation into Z boson}

First of all, we show our results for the semi-annihilating channel $DM + DM \rightarrow DM + Z$, where $Z$ is the SM Z boson. In this case, in order to see the impact of the dark matter mass in the gamma-ray spectrum, we choose the dark matter mass to be equal to $91$~GeV (solid line), $1000$~GeV (dashed line) and $10000$~GeV (dotted line) as displayed in the left-panel of Figure~\ref{semiz}. As expected the higher the dark matter mass the harder the gamma-ray spectrum. The shape of the spectrum remains the same though. \\

This information allows us to understand the sensitivity of each experiment. The Fermi-LAT experiment is more sensitive to gamma-rays at low energies, whereas H.E.S.S and CTA to larger energies. The energy threshold of the CTA is lower than the one present in H.E.S.S., making the CTA much more sensitive to gamma-rays at intermediate energies, and consequently at $100$~GeV dark matter masses. Bearing in mind this information, one can understand why Fermi-LAT provides much stronger bounds on the semi-annihilation cross section for dark matter masses below $100$~GeV.\\

\begin{figure}[ht]
\centering
\includegraphics[width=0.45\columnwidth]{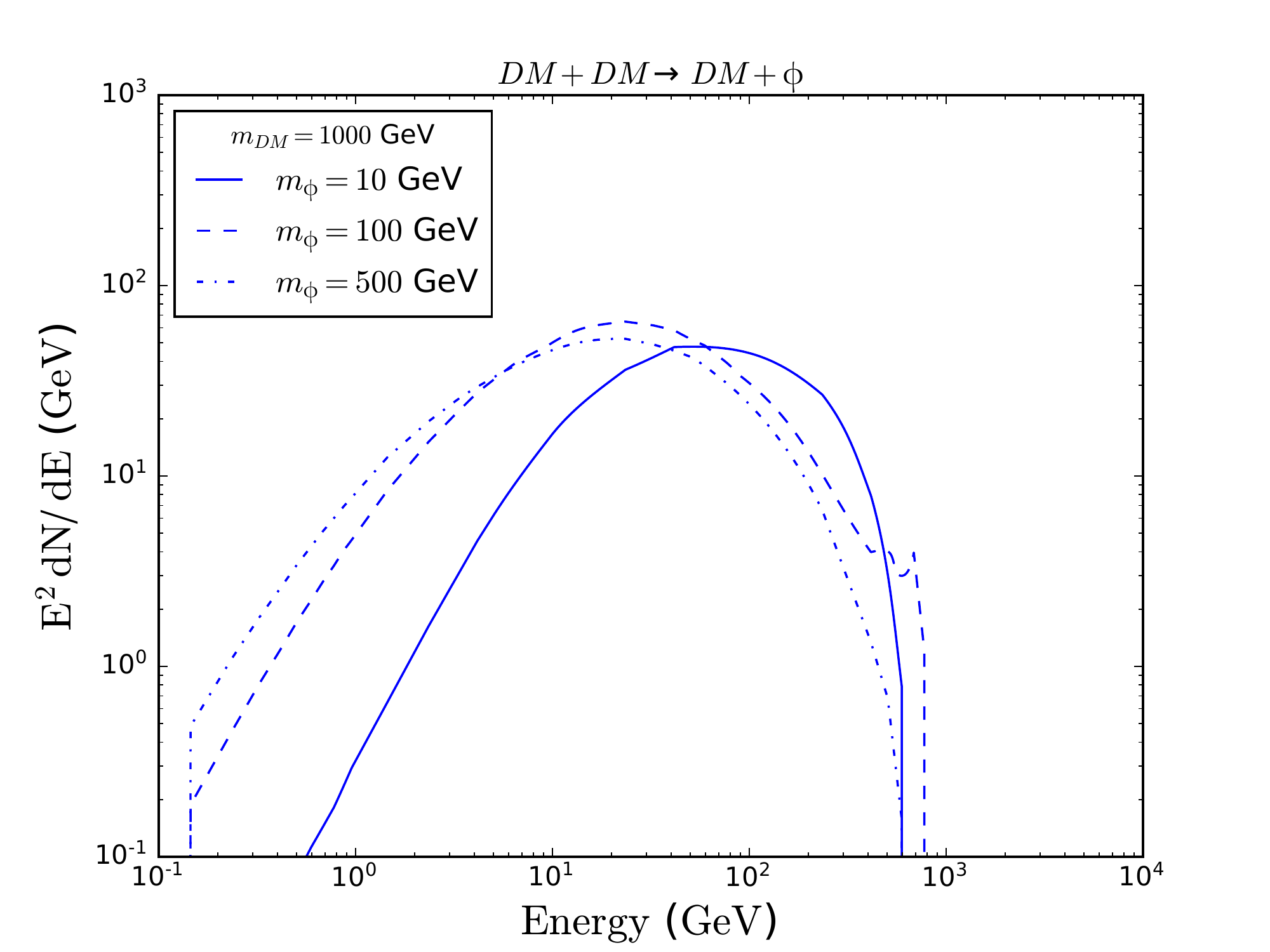}
\includegraphics[width=0.45\columnwidth]{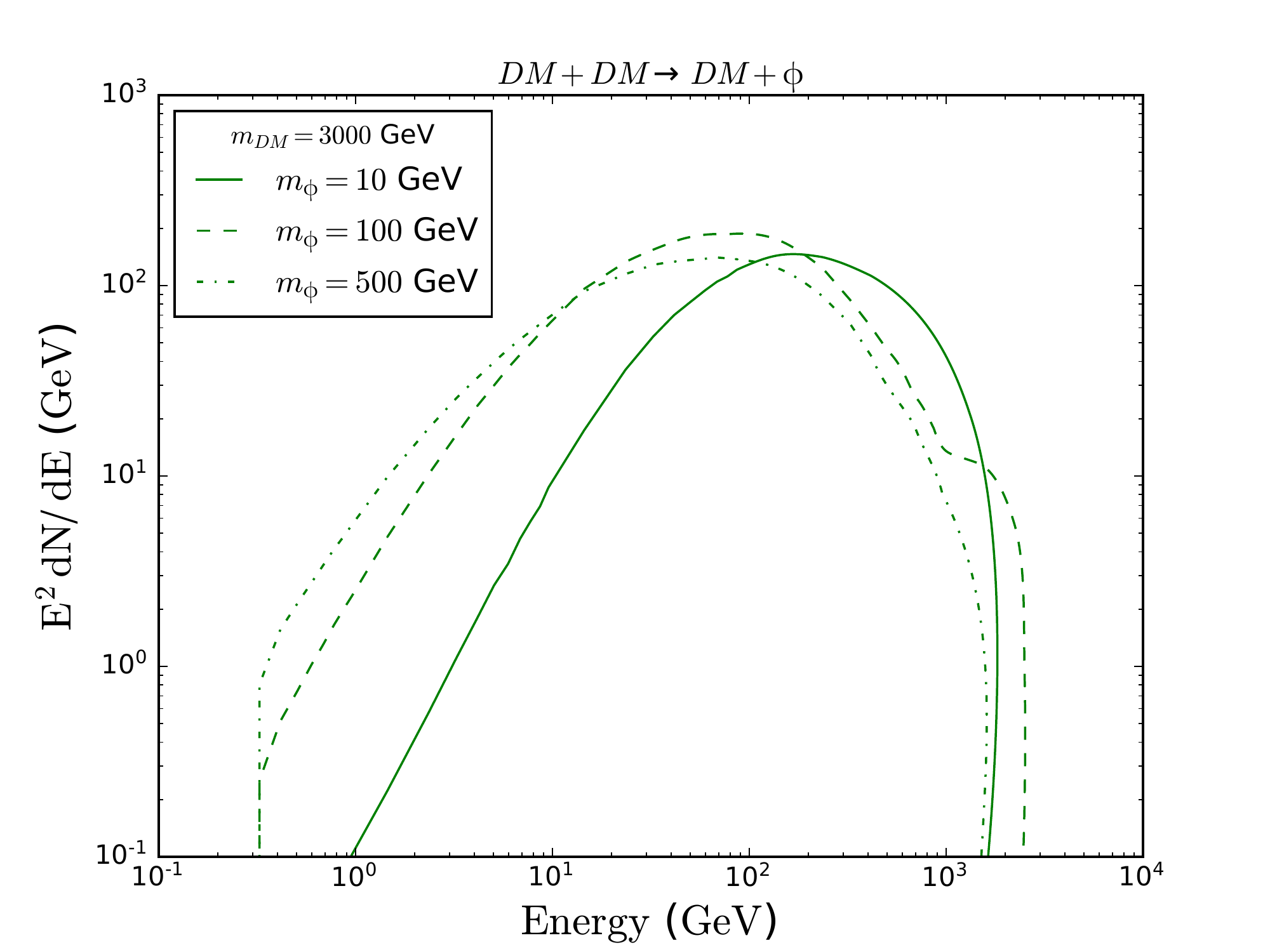}
\caption{Gamma-ray spectrum for DM annihilating in the Semi-$\phi$ (Higgs-like particle) channel, in the left panel, we fix the DM mass equal to $1000$~GeV and in the right panel equal to $3000$~GeV, in both cases varying the $\phi$ mass equal to $10$~GeV (continuous lines), $100$~GeV (dashed lines) and $500$~GeV (dotted lines), as high as the DM mass is more energetic gamma-rays will be produced, the shape related to the $\phi$ masses depends on the branching ratio.}
\label{spectrasemiphihiggslike}
\end{figure}

There are additional factors that go into the upper limits on the semi-annihilation cross-section shown in the {\it right-panel} of Figure~\ref{semiz}. This figure refers to the channel $DM + DM \rightarrow DM + Z$. The large branching ratio of the Z boson into hadrons explains the poor CMB bounds. The CMB observable is related to electromagnetic energy injection which is less efficient when the decay channel is hadronic. This goes in the opposite direction of gamma-ray telescopes. For this reason, for $m_{DM}< 1 $~TeV Fermi-LAT reports the best limits, while H.E.S.S. is more restrictive for $m_{DM} > 1$~TeV. The projected CTA sensitivity to the semi-annihilation into $DM, \, Z$ is very important to test this scenario since it will yield the most stringent bounds for $m_{DM} > 200$~GeV.

\subsection{Semi-Annihilation into Higgs}

For the semi-annihilation case $DM +DM \rightarrow DM + h$, where $h$ is the SM Higgs boson, the gamma-ray spectrum follows the same behavior as for the Z boson, dominated by hadronic channels. In the {\it left-panel} of Figure \ref{semih} we display the energy spectrum for $m_{DM}=125,\,1000,\,10000$~GeV, which exhibits a continuum emission and a peak at the end of the spectra due to the branching fraction of the Higgs in $\gamma \gamma$ \cite{Profumo:2016idl}. \\

The results are quite similar to the $DM + DM \rightarrow DM Z$ scenario with gamma-ray telescopes rendering the strongest bounds.

\subsection{Semi-Annihilation into Higg-Like Particles}

\begin{figure}[ht]
\centering
\includegraphics[width=0.49\columnwidth]{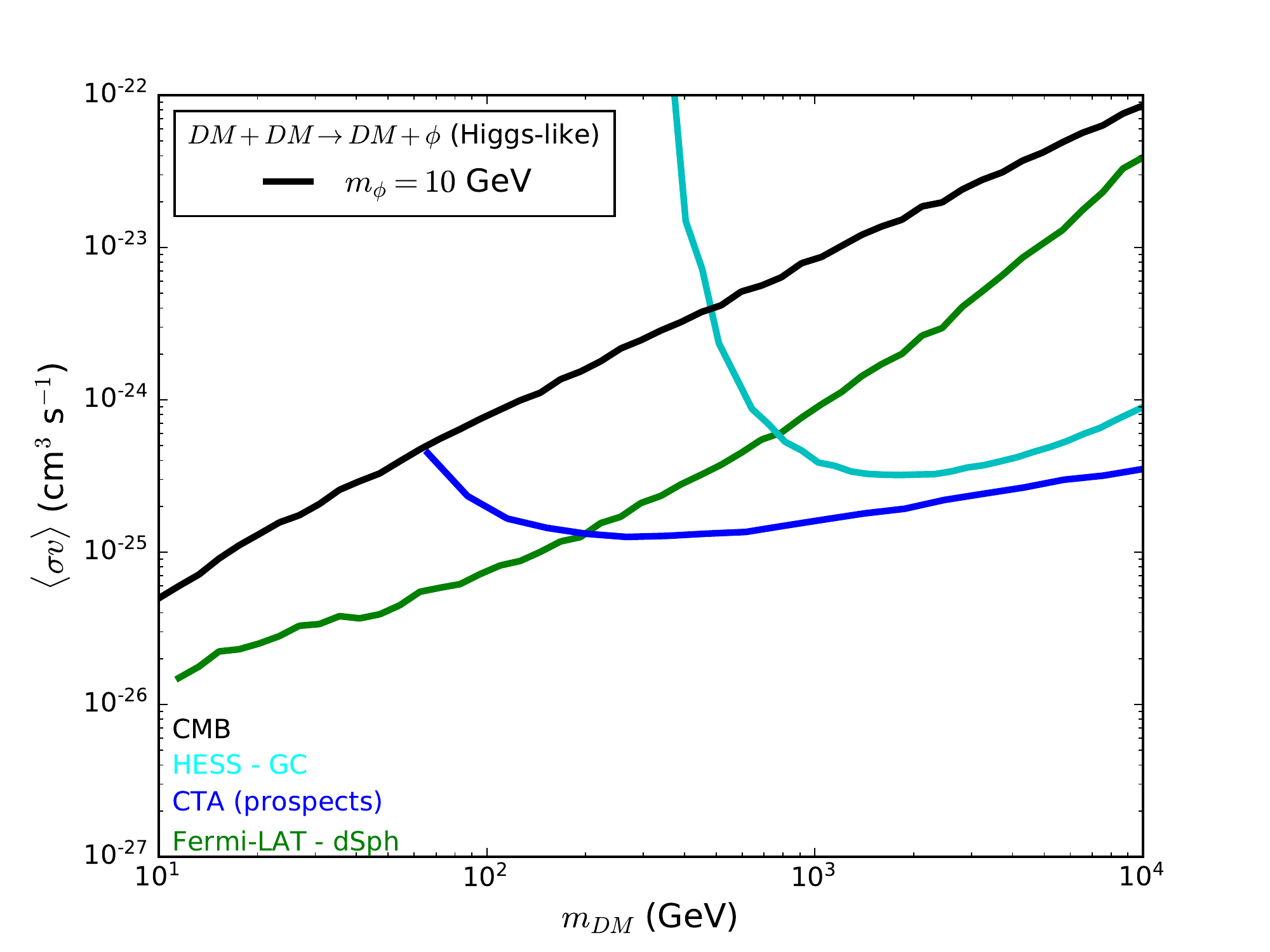}
\includegraphics[width=0.49\columnwidth]{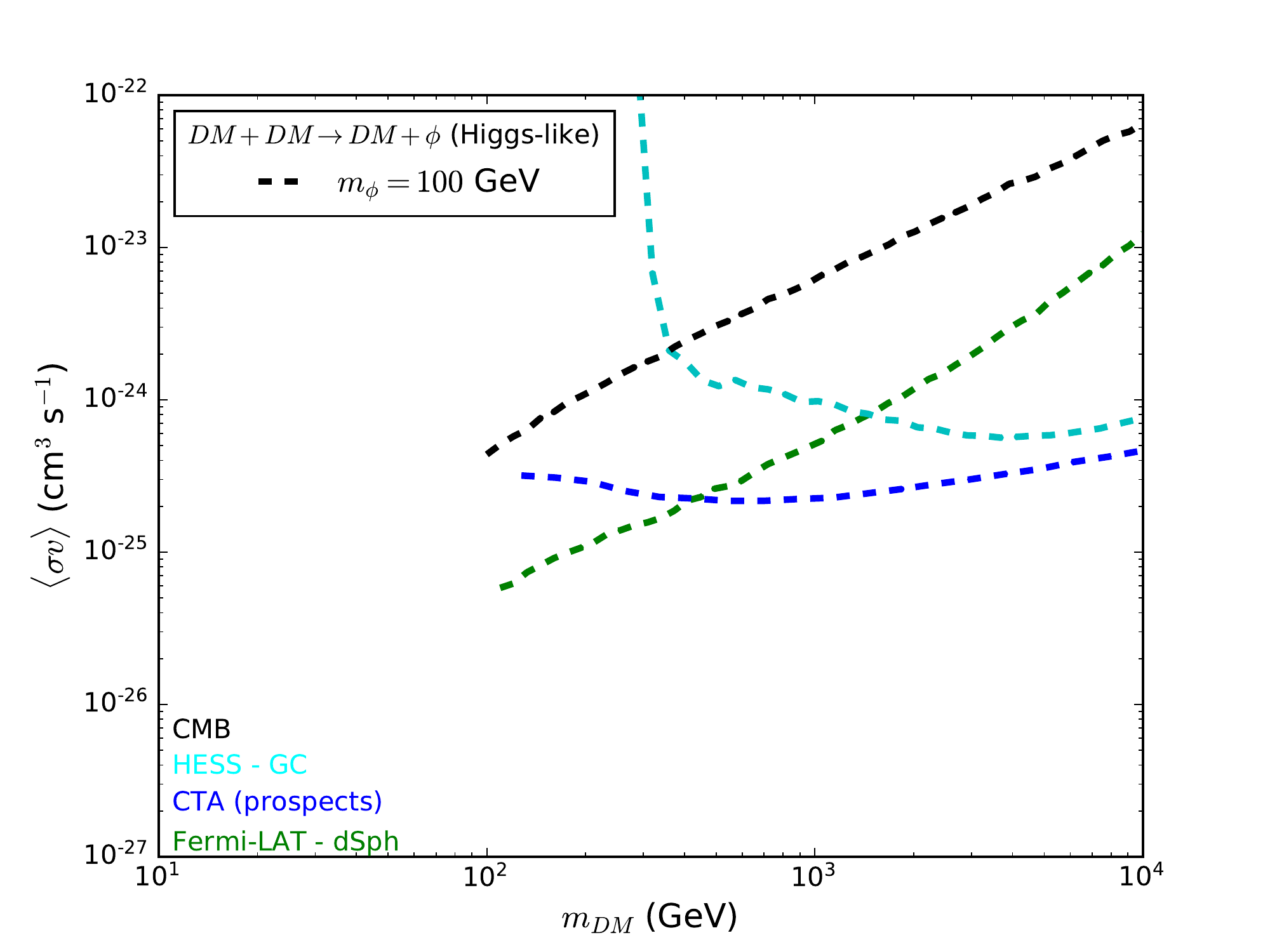}
\includegraphics[width=0.49\columnwidth]{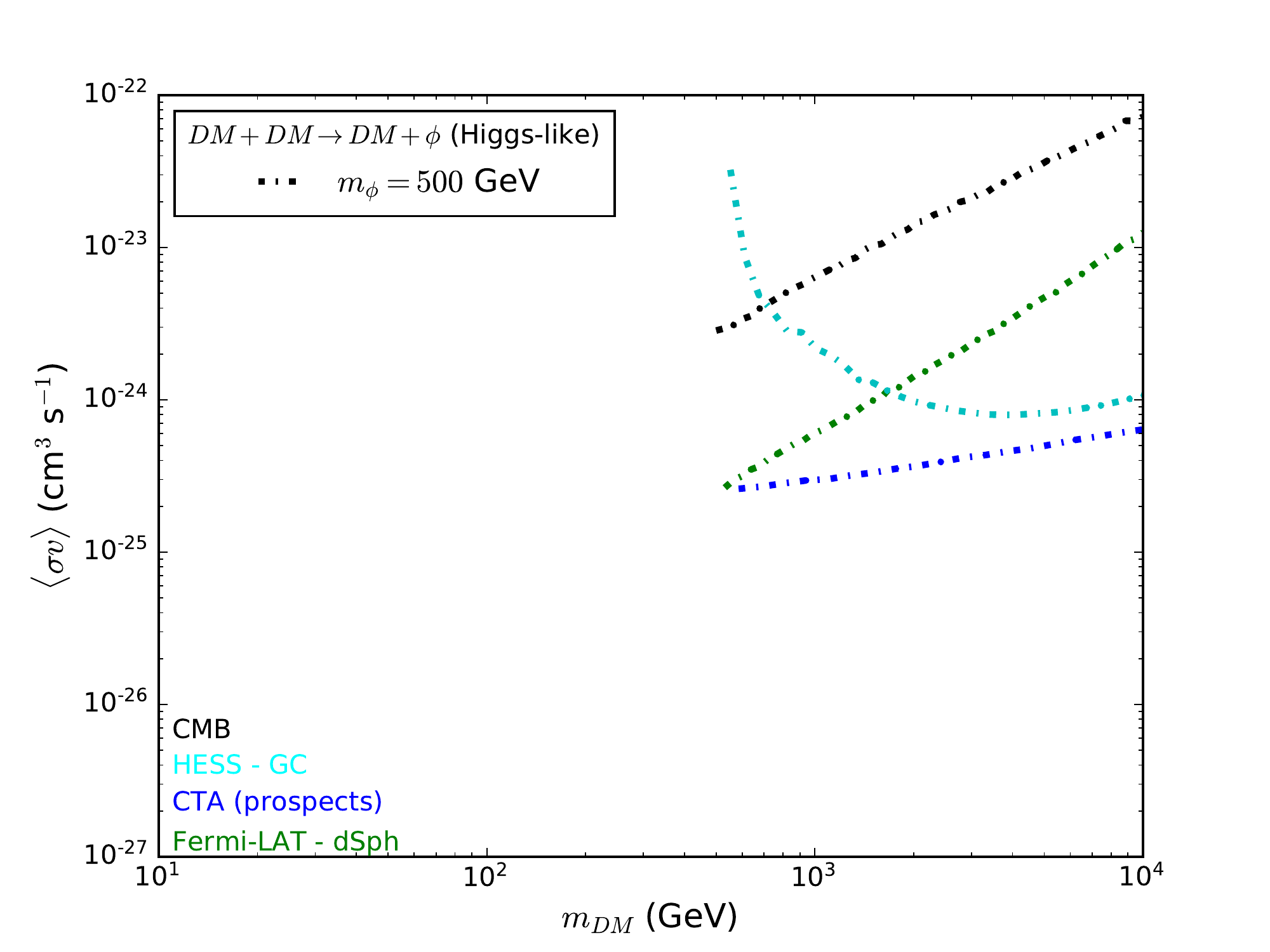}
\caption{$95\%$~C.L. upper limits on the DM annihilation cross section for the channel $DM+DM \rightarrow DM + \phi$ (Higgs like), choosing the mass of the $\phi$ particle equal to $10,100,500$~GeV for each experiment: Planck (black), H.E.S.S. (cyan), Fermi-LAT (green), CTA prospect (blue).}
\label{semiphihiggslike10}
\end{figure}

Now, we will present experimental limits on the semi-annihilation into Higgs-like particles, $DM +DM \rightarrow DM + \phi$, where $\phi$ is a Higgs-like scalar field with a different mass. This scalar couples to all fermions, in the same way, the SM Higgs. The only difference is its own mass which we vary. In Figure~\ref{spectrasemiphihiggslike}, we show the gamma-ray spectra for $m_{DM}=1000$~GeV (left-panel) and $m_{DM}=3000$~GeV (right-panel). In each case, we provide the energy spectra for three different values masses of the Higgs-like particle, $m_{\phi}=10$~GeV (continuous lines), $m_{\phi}=100$~GeV (dashed lines) and $m_{\phi}=500$~GeV (dotted lines). As we expected, the spectra is harder for larger dark matter masses, and the change on the spectra due to the $\phi$ particle mass depends on the branching ratio, for example, the peak due to the branching ratio in $\gamma \gamma$ is visible when $m_\phi=100$~GeV, but not in the other cases, this is due to the suppressed branching in $\gamma \gamma$. \\


In the Figure~\ref{semiphihiggslike10}, we present the upper limits in the annihilation cross section versus the dark matter mass for the $\phi$ mass equal to $10$~GeV, $100$~GeV and $500$~GeV, respectively. The behavior of the curves follows the same pattern like in the standard Higgs case, with the CMB bound not being competitive with the other indirect searches. In addition, increasing the mass of the scalar field $\phi$, for instance, for the $\phi$ mass equal to $100$~GeV and $500$~GeV, we can see a cut in the limits in the dark matter mass, this is due to energy conservation. For the same reason we can notice that for light $\phi$, the limits extend to light dark matter masses as well. Once again, CTA will yield the best gamma-ray limits for dark matter masses above the few hundred GeV.

\subsection{Semi-Annihilation into Leptophilic and Leptophobic Scalars}

Our last analysis concerns semi-annihilation into leptophilic scalars, $DM + DM \rightarrow DM + \phi$, where $\phi$ is a leptophilic scalar decaying $100\%$ into $e^+e^-$, $\mu^+\mu^-$, or $\tau^+\tau^-$, or leptophobic with $\phi$ decaying exclusively into $b\bar{b}$. For each channel, we vary the mass of the scalar in order to see the impact in the results.  

\begin{figure}[ht]
\centering
\includegraphics[width=0.49\columnwidth]{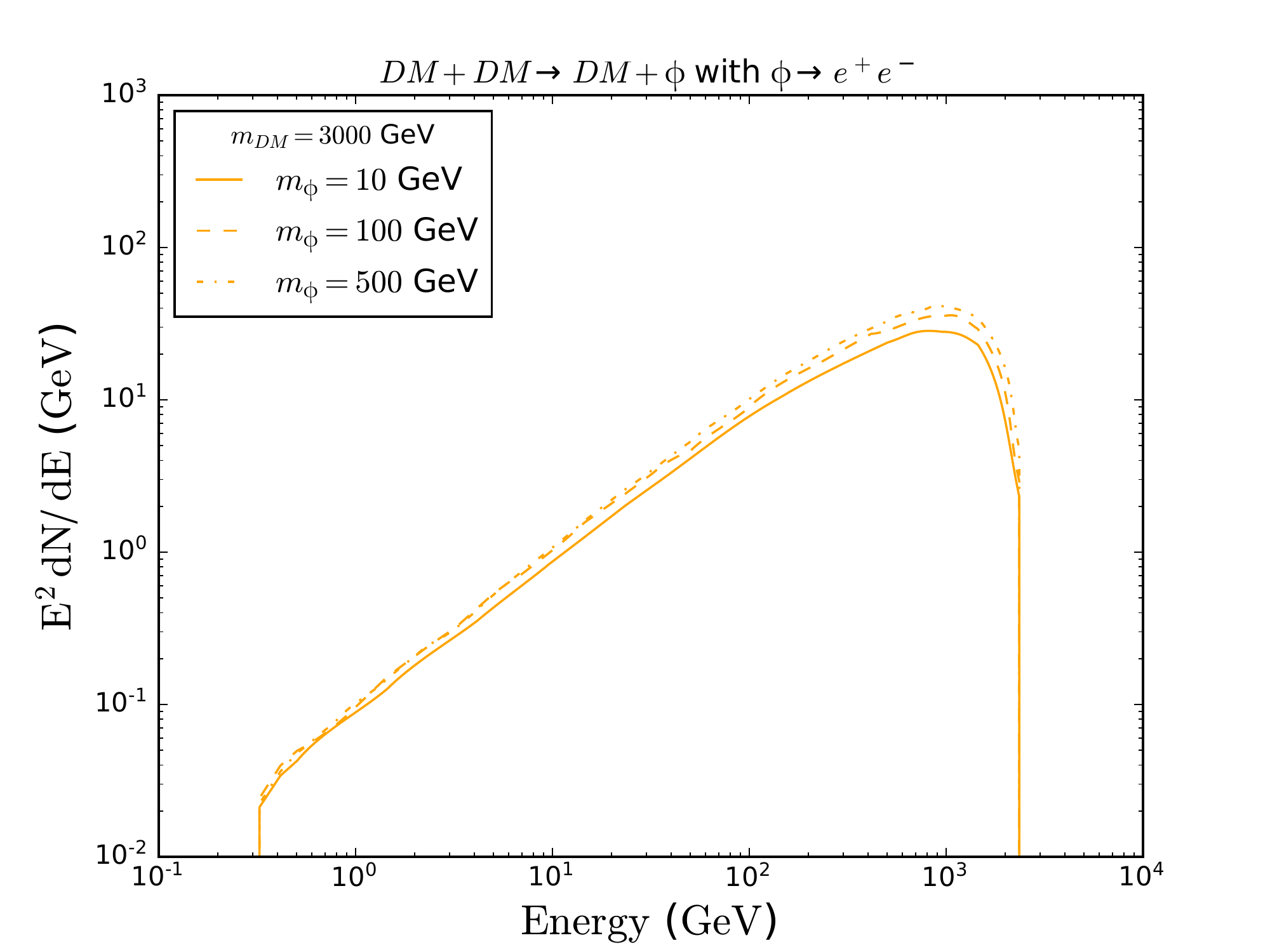}
\includegraphics[width=0.49\columnwidth]{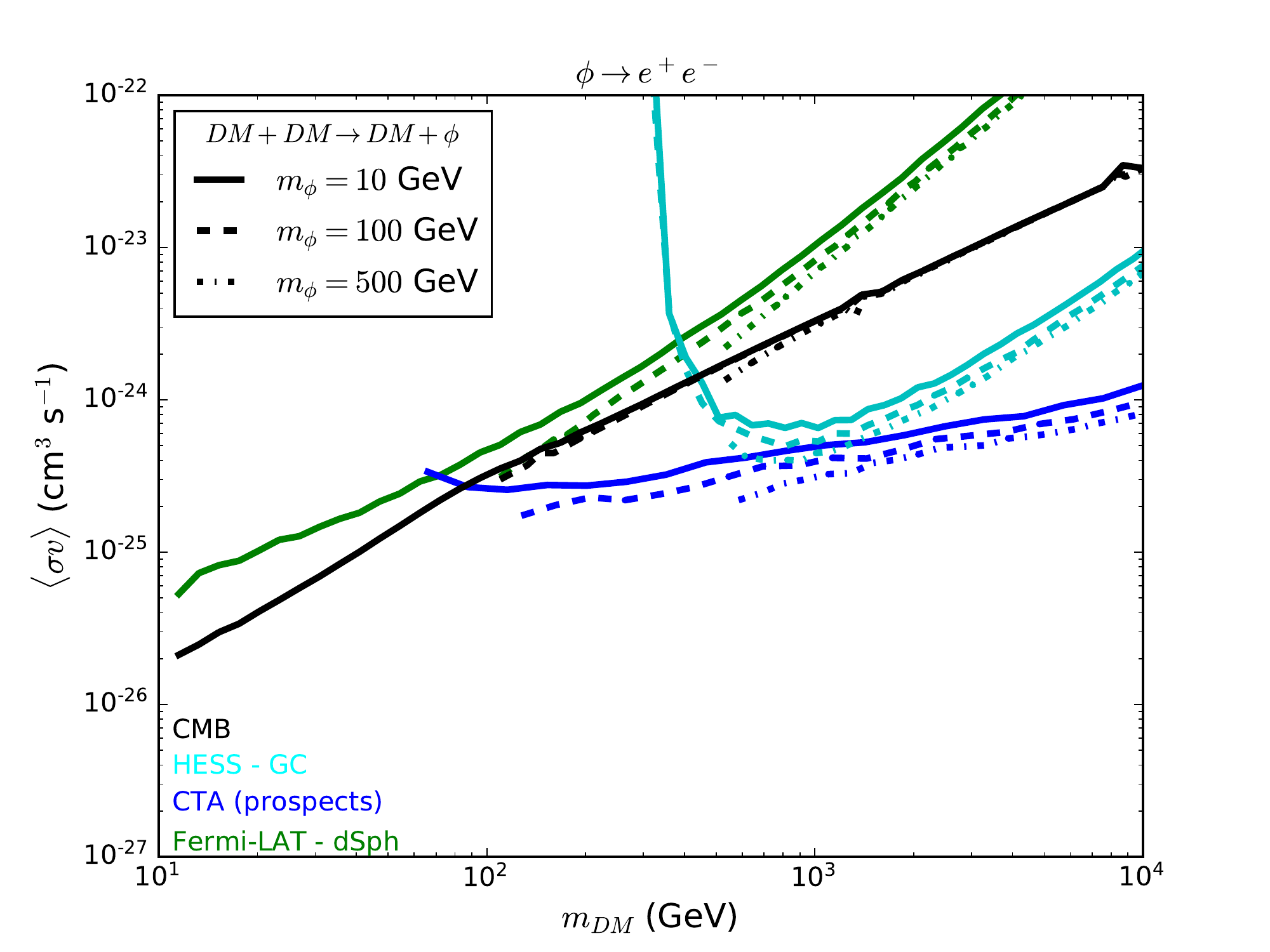}
\caption{\textbf{Left-panel}: Gamma-ray spectrum for dark matter semi-annihilation, $DM+ DM \rightarrow DM + \phi$, with $\phi \rightarrow e^+ e^-$. We are fixing the dark matter mass at $3000$~GeV and varying the scalar mass in $10$~GeV (continuous lines), $100$~GeV (dashed lines) and $500$~GeV (dotted lines).. \textbf{Right-panel}: Limits from different experiments over the semi-$\phi$ channel. Black lines for Planck constraints, cyan curves for the H.E.S.S. bounds, green lines for the Fermi-LAT exclusion curves, and blue lines for the CTA prospects.}
\label{semi100ee}
\end{figure}

\begin{figure}[ht]
\centering
\includegraphics[width=0.49\columnwidth]{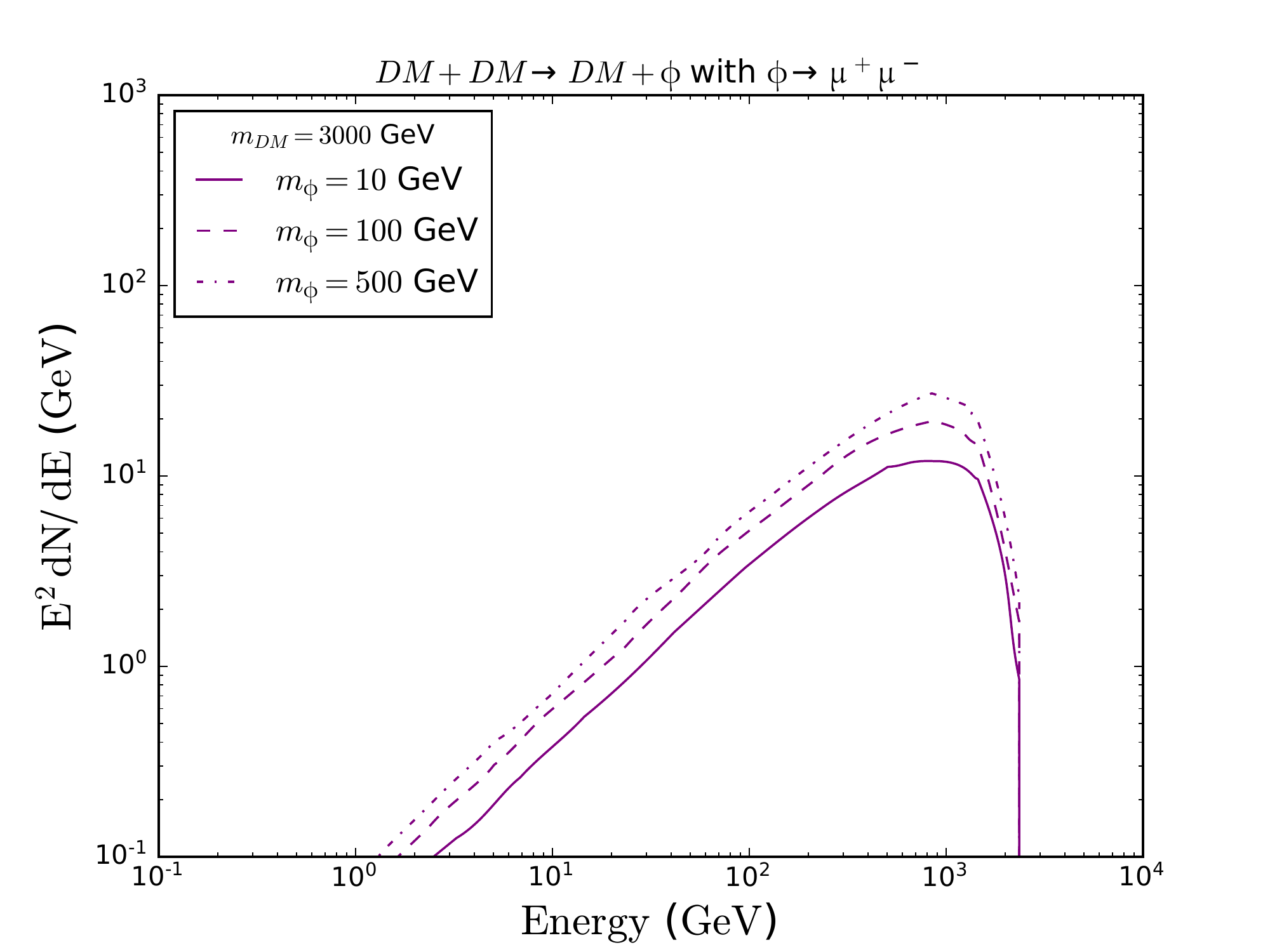}
\includegraphics[width=0.49\columnwidth]{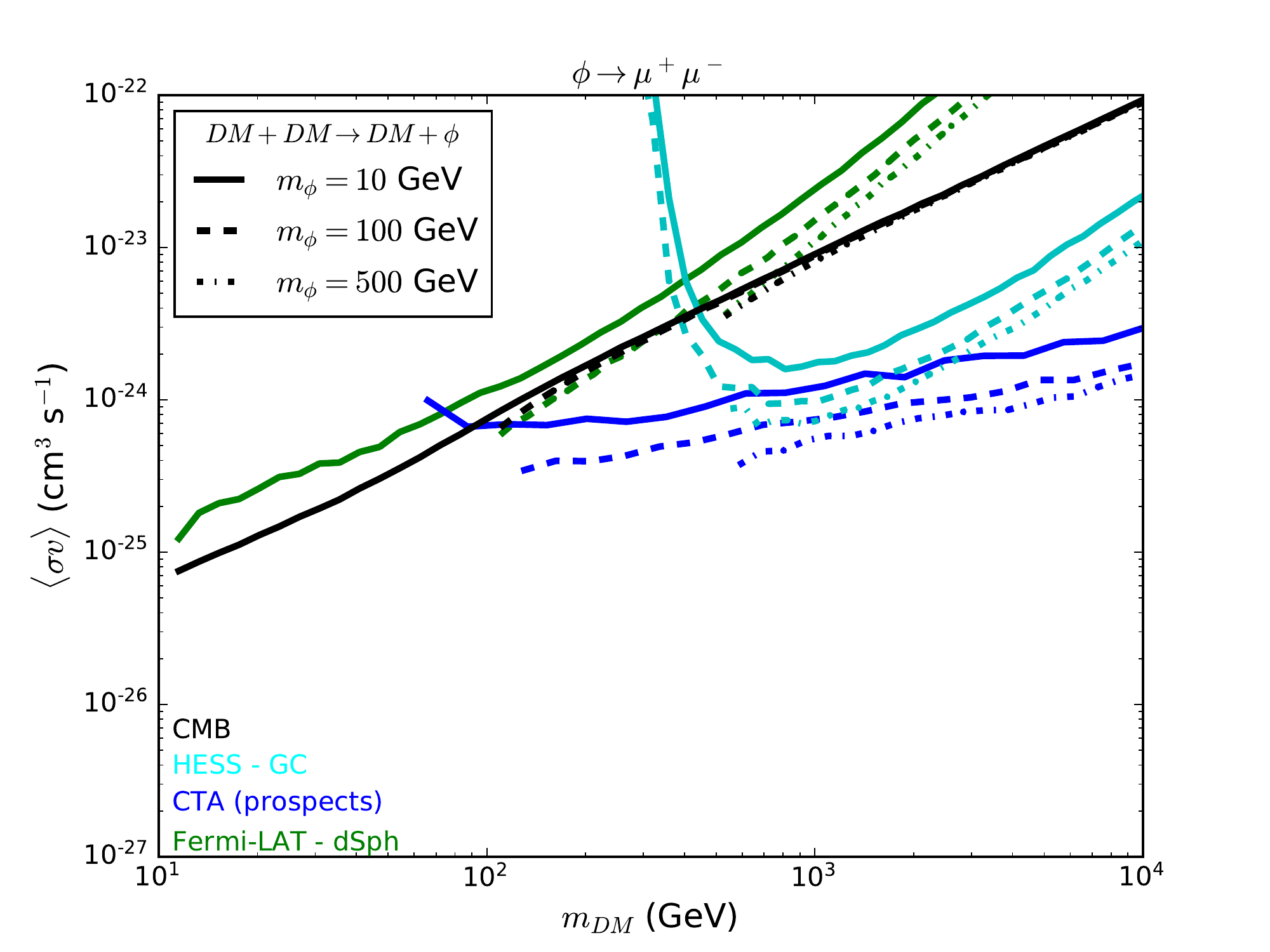}
\caption{\textbf{Left-panel}: Gamma-ray spectra for dark matter semi-annihilation, $DM+ DM \rightarrow DM + \phi$, with $\phi \rightarrow \mu^+ \mu^-$. We are fixing the dark matter mass at $3000$~GeV and varying the scalar mass in $10$~GeV (continuous lines), $100$~GeV (dashed lines) and $500$~GeV (dotted lines). \textbf{Right-panel}: Upper limits from different experiments on the semi-annihilation cross section for different setups from Planck (black line), H.E.S.S. (cyan) and Fermi-LAT (green), and CTA (blue).}
\label{semi100mumu}
\end{figure}

\begin{figure}[ht]
\centering
\includegraphics[width=0.45\columnwidth]{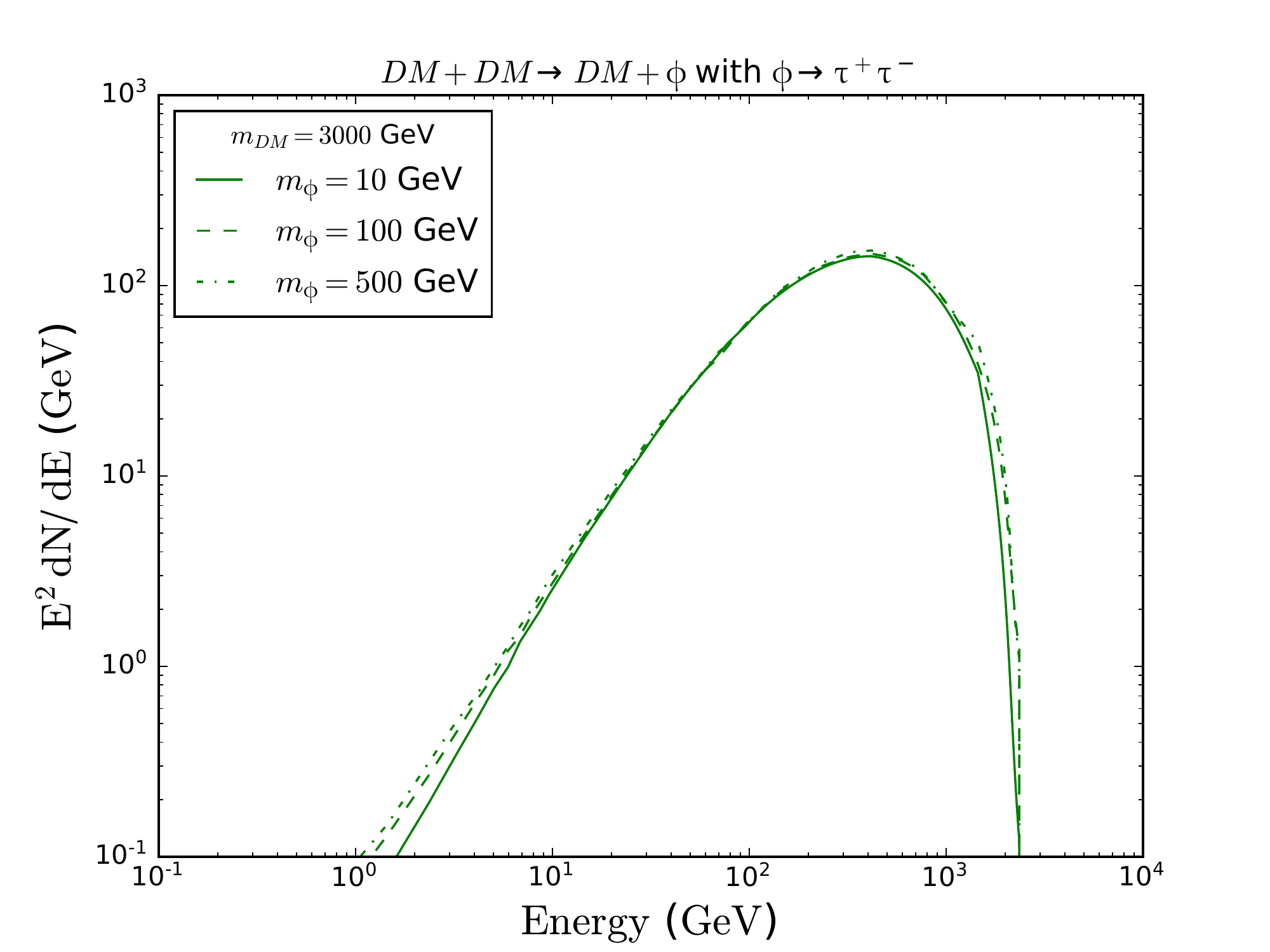}
\includegraphics[width=0.45\columnwidth]{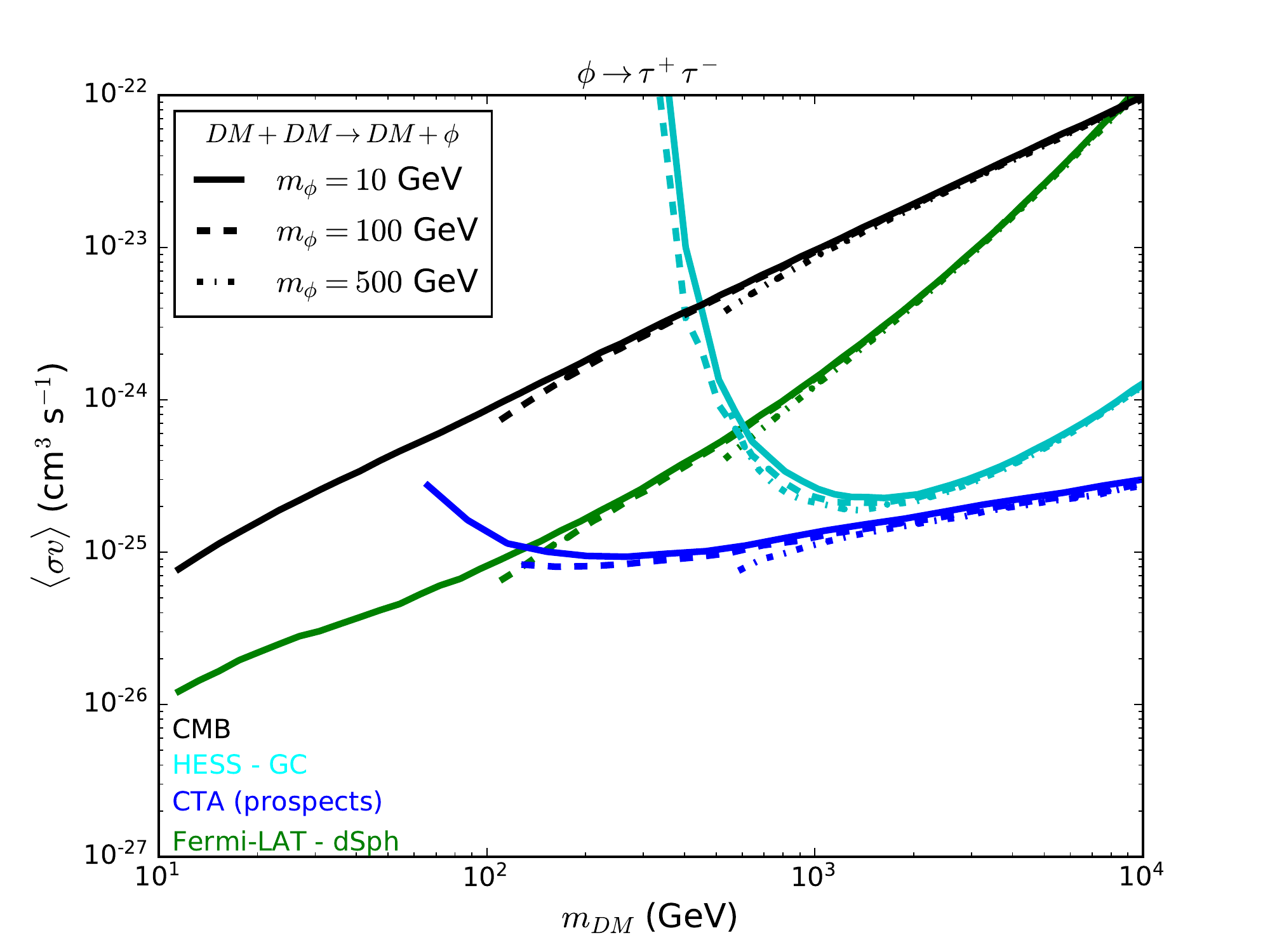}
\caption{\textbf{Left-panel}: Gamma-ray spectrum for DM annihilating in the Semi-$\phi$ channel with $\phi$ decaying predominantly in $\tau^+ \tau^-$, here we are fixing the DM mass at $3000$~GeV and varying the scalar mass in $10$~GeV (continuous lines), $100$~GeV (dashed lines) and $500$~GeV (dotted lines). \textbf{Right-panel}:  Upper limits from different experiments on the semi-annihilation cross section for different setups from Planck (black line), H.E.S.S. (cyan) and Fermi-LAT (green), and CTA (blue).}
\label{semi100tautau}
\end{figure}

\begin{figure}[ht]
\centering
\includegraphics[width=0.49\columnwidth]{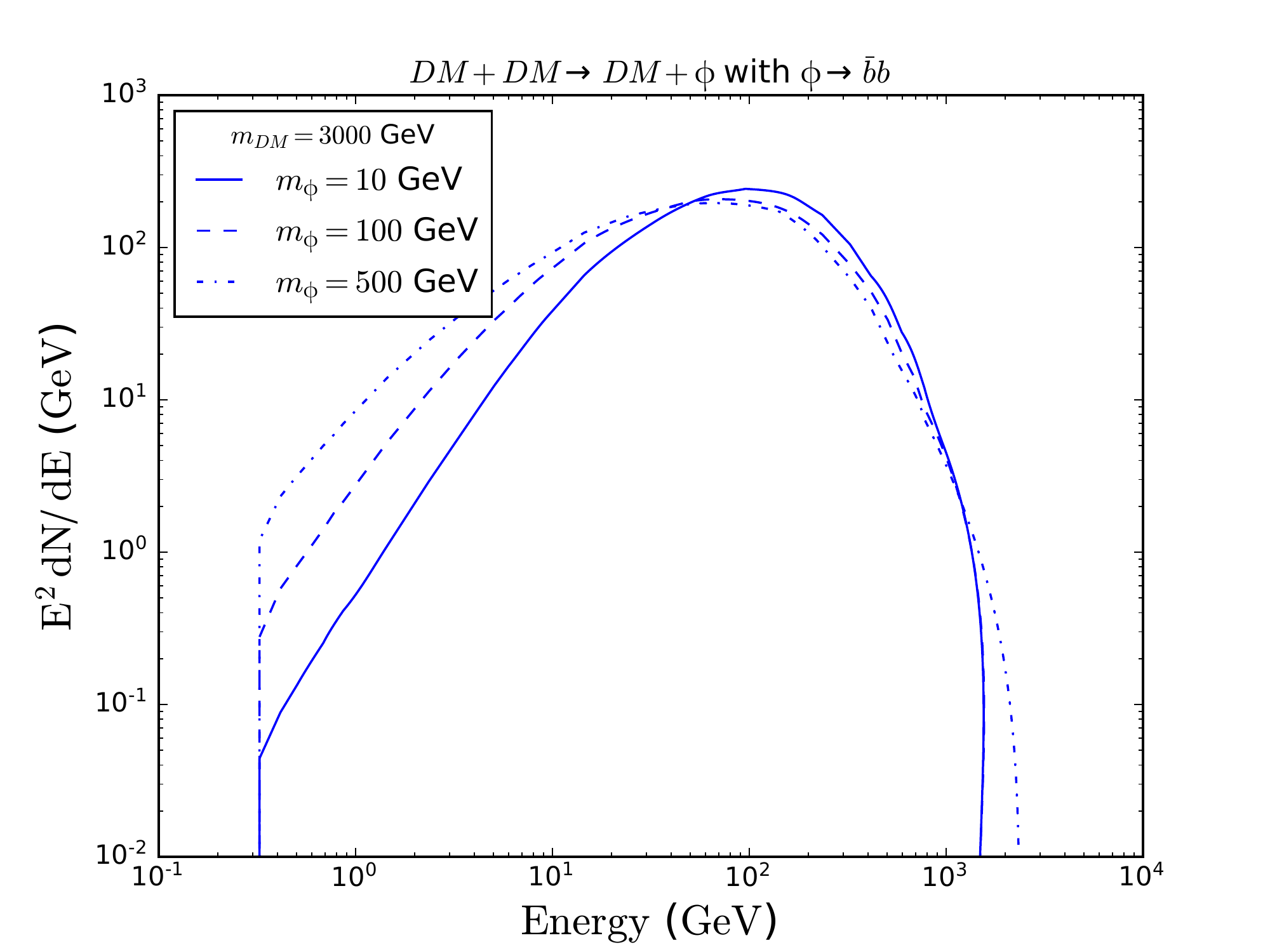}
\includegraphics[width=0.49\columnwidth]{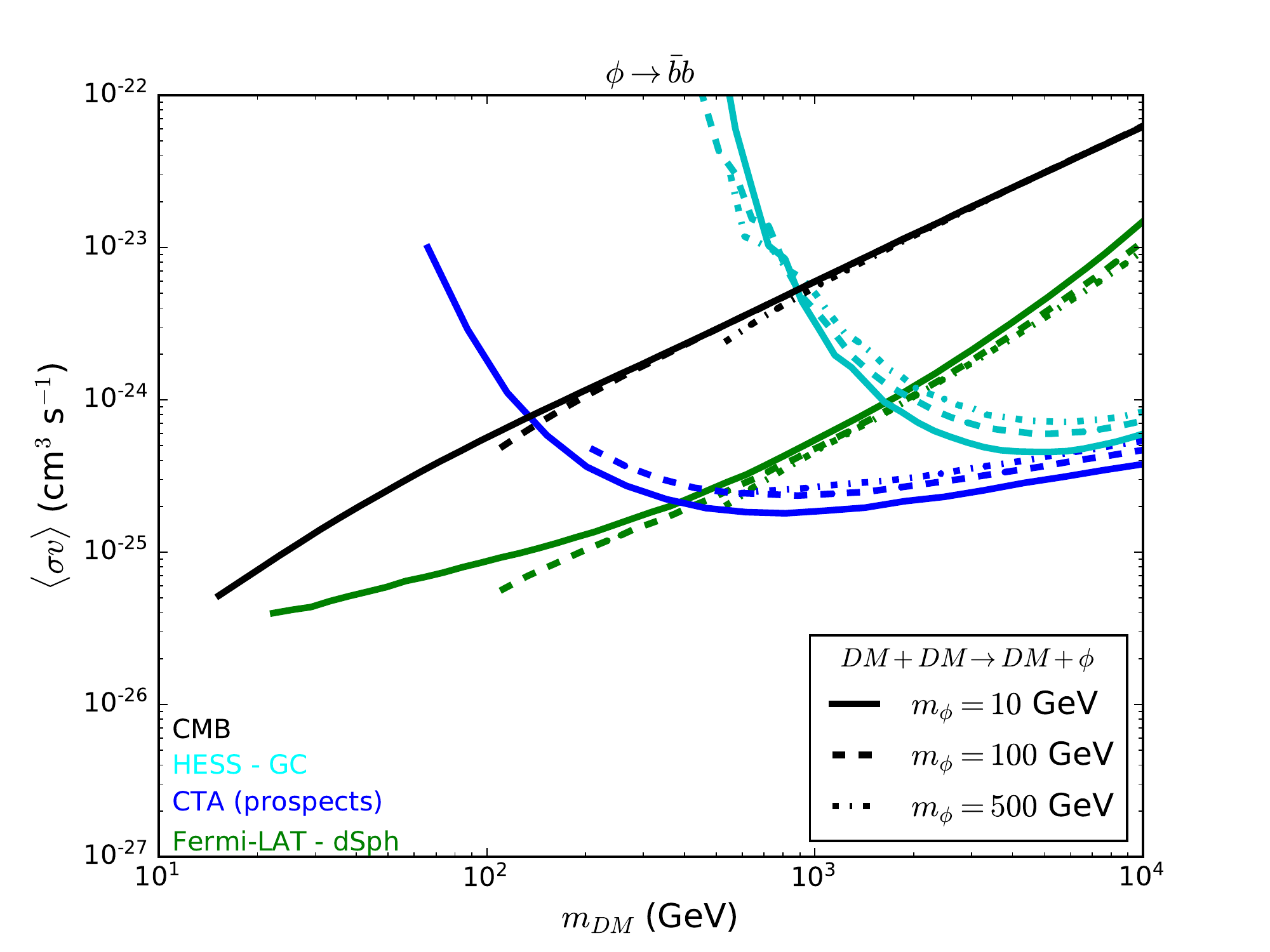}
\caption{\textbf{Left-panel}: Gamma-ray spectrum for DM annihilating in the Semi-$\phi$ channel with $\phi$ decaying predominantly in $\bar{b}b$, here we are fixing the DM mass at $3000$~GeV and varying the scalar mass in $10$~GeV (continuous lines), $100$~GeV (dashed lines) and $500$~GeV (dotted lines). \textbf{Right-panel}: Upper limits from different experiments on the semi-annihilation cross section for different setups from Planck (black line), H.E.S.S. (cyan) and Fermi-LAT (green), and CTA (blue).}
\label{semi100bb}
\end{figure}

In the Fig,~\ref{semi100ee}, we show our results for the case which $\phi$ decays only into $e^+ + e^-$.  In the {\it left-panel}, display the gamma-ray spectrum fixing the dark mass in $3000$~GeV and varying the $\phi$ mass from $10$~GeV (continuous lines), $100$~GeV (dashed lines), up to $500$~GeV (dotted lines). The larger the $\phi$ mass the harder the gamma-ray spectrum. In this setup, the difference is very modest though. This reflects directly in the upper bound on the semi-annihilation cross section as shown in the {right-panel} of Figure~\ref{semi100ee}. In a similar vein, we show the results for predominant decays into muons in Figure~\ref{semi100mumu}. Notice that CMB limits are rather strong for these channels because of the large electromagnetic energy injection.\\

It is typically assumed for dark matter annihilations into charged leptons CMB constitutes the best probe. If we take a look at the {\it right-panel} of  Figure~\ref{semi100mumu}, in particular the result for $m_\phi=10$~GeV, we notice that Planck limits are indeed the strongest ones for $m_{DM}< 100$~GeV. However, taking $m_\phi=100$~GeV, in the mass region $100\,{\rm GeV} <m_{DM} < 400$~GeV, the gamma-ray limits from Fermi-LAT as stringent as the ones from Planck. Furthermore, for $m_{DM}> 400$~GeV, the limits from H.E.S.S. are much stronger than those based on CMB measurements. Therefore, this naive assumption that CMB bounds are the most restrictive annihilations into charged leptons does not necessarily apply to semi-annihilations. This highlights the importance of considering orthogonal indirect detection probes. \\

For semi-annihilations into scalars, where the scalar decays into taus, we find the upper limits shown in Figure~\ref{semi100tautau}. The CMB bounds weaken because of the hadronic decays that produce lots of gamma-rays via pion production. The upper bounds from H.E.S.S. exclude  $\sim 10^{-25}$~cm$^3/$s for $m_{DM}\sim 1$~TeV. The prospects for CTA are quite amazing regardless of the final state involved and probe much lower masses compared to H.E.S.S. due to the different set of telescopes adopted capable of measuring gamma-rays at much lower energies \cite{Acharya:2017ttl}\\

Lastly, in Figure~\ref{semi100bb}, we show our upper limits on the semi-annihilation $DM+DM \rightarrow DM+\phi$ with $\phi \rightarrow \bar{b} + b$, in the {\it left-panel}, the gamma-ray spectrum fixing the dark matter mass in $3000$~GeV and varying the $\phi$ mass in $10$~GeV (continuous lines), $100$~GeV (dashed lines) and $500$~GeV (dotted lines). In this case, the production of $\bar{b}b$ which quickly hadronize producing charged and neutral pions whose decay yield lots of gamma-rays making the results stronger. In this scenario, it is visible the impact of changing the $\phi$ mass for low energy photons. The energy spectrum for $m_{\phi}=10$~GeV is softer than the one for $m_{\phi}=500$~GeV. This is visible in the {\it right-panel} of Figure~\ref{semi100bb} where one can notice that the upper bound on the semi-annihilation cross section for $m_{\phi}=500$~GeV (dotted curve) is always below the one for $m_{\phi}=10$~GeV (solid line). On the other hand, for $m_{\phi}=10$~GeV, one can place limits on the semi-annihilation cross section down to $\sim 10$~GeV. The CMB limits for this hadronic channel becomes deadened as explained earlier.\\ 

Anyway, it is clear from our results that in order to solidly probe semi-annihilating dark matter, one needs to go beyond one dataset. The upper limits obtained from Fermi-LAT, Planck and H.E.S.S. data are quite complementary to one another, covering different mass regions and semi-annihilation modes.

\section{Conclusions}
\label{sec:con}

In most dark matter models, the parameters that govern the self-annihilation processes are the same that dictate the scattering rates at underground laboratories. Direct detection experiment has placed stringent bounds on the dark matter scattering cross section, disfavoring several dark matter models. One way to alleviate the tension between the null results from direct detection is to invoke semi-annihilation, which can be quite relevant. In this way, one can potentially break the degeneracy between direct detection and dark matter self-annihilation. That said, in the absence of experimental limits on semi-annihilations we used different datasets stemming from gamma-ray observations from Dwarf Spheroidal Galaxies, the Galactic Center and CMB to derive the first upper bounds on the dark matter semi-annihilation cross section for semi-annihilations into Z, Higgs, Higgs-like, Leptophilic and Leptophobic scalars over a wide range of masses. Moreover, we presented the CTA prospects for each case and explored the complementary aspects of each experiment and dataset. Our findings represent the strongest bounds on the semi-annihilation cross section, excluding $\sigma v \sim 3 \times 10^{-26}$~cm$^3/$s for $m_{DM} = 20$~GeV, for semi-annihilation into leptophobic scalars, for instance.

\section*{Acknowledgments}

We would like to thank the Fermi-LAT Collaboration for the public data. We are grateful to Aion Viana, Vitor de Souza, and Carlos Yaguna for discussions. We are also thankful to Manuela Vecchi for reading the manuscript and suggestions. This work was supported by MEC, UFRN and ICTP-SAIFR FAPESP grant 2016/01343-7. We thank the High Performance Computing Center (NPAD) at UFRN for providing computational resources.

\bibliographystyle{JHEPfixed}
\bibliography{darkmatter}

\end{document}